\documentstyle[12pt]{article}
\newcommand{\be}{\begin{equation}}
\newcommand{\ee}{\end{equation}}
\newcommand{\ba}{\begin{eqnarray}}
\newcommand{\ea}{\end{eqnarray}}
\newcommand{\ds}{\displaystyle}

\begin{document}

\begin{titlepage}

\begin{center}
{\large\bf Neutrino pair production by a virtual photon\\
in an external magnetic field}\\
\end{center}
\vspace{0.5cm}
\begin{center}
V.Ch. Zhukovskii, A.E.Grigoruk, K. G. Levtchenko \\
{\sl
Faculty of Physics,\\
Department of Theoretical Physics, Moscow State University,\\
119899, Moscow, Russia} \\
P. A. Eminov\\
{\sl
Department of Physics\\
Moscow State Institute of Electronics and Mathematics\\
(Technical University)\\
109028, Moscow, Russia}\\[15pt]
\end{center}

\begin{abstract}
Production of a massive neutrino--antineutrino pair by a
virtual polarized photon in the Weiberg--Salam model with mixing is
studied. The rate of the neutrino production by photons with
various polarizations is found at values of the magnetic field
intensity lower than critical.
\end{abstract}

\vspace{0.5cm}

\vfill
\end{titlepage}

     Investigation of various production and decay mechanisms of a
massive neutrino is of principle interest for the test of the
Standard Model of elementary particle interactions and its possible
extensions, as well for its astrophysical applications [1--4]. In
the present paper the decay of a polarized virtual photon in a
constant magnetic field into the neutrino--antineutrino pair in the
framework of the Standard Model with lepton mixing is studied.

     The main contribution to the matrix element of the process
$\gamma^* \to \bar \nu_i \nu_j$ in the Feynman gauge in the contact
approximation can be written in the following form (for more
details see [2, 4]):
\ba \label{tot}
M^{tot}_{\gamma^* \to \bar \nu \nu}&=&-i{\ds\frac{4eG_F}{\sqrt{2}}}
\varepsilon^\alpha(k)j^\beta \langle K^a_{\alpha\beta} \rangle_a,\\
\langle K^a_{\alpha\beta} \rangle_a&=&( {\ds\sum_a} U_{aj}^*U_{ai} + 2
\delta_{ij} )  {\ds\int \frac{d^4p}{(2\pi)^4}}\,
\mbox{Tr}\,[\gamma_\beta(1+\gamma^5)S^a(p)\gamma_\alpha
S^a(p+k)].\nonumber
\ea
Here $S^a(p)$ is the charged lepton propagator $(a=e, \mu, \tau)$
in an external field [5],
$\ds j^\beta=\overline{\nu}_j(q')
\gamma^\beta \frac{(1+\gamma^5)}{2}\nu_i(-q)$ is the ``neutrino current'',
$\overline{\nu}_j(q')$ is the Dirac conjugate bispinor of the final
neutrino with the momentum $q'$, $\nu_i(-q)$ is the bispinor of the final
antineutrino with the momentum $-q$ $(q'^2 = m_j^2, q^2 = m_i^2)$.

     Next, being interested in the process rate only as a function
of an external field, the case of a comparatively weak magnetic
field with the intensity $H \ll H_0^a \equiv m_a^2/e$ is
considered, and the amplitude (1) is
separated into the parts that correspond to three independent
polarization states of the decaying virtual photon, that are
described by the following 4--vectors of polarization:
\ba
\varepsilon^{(1)}_\alpha&=&\frac{(Fk)_\alpha}{\sqrt a},\,\,\,
\varepsilon^{(2)}_\alpha=\frac{(\tilde F k)_\alpha}{\sqrt \beta},\nonumber\\
\varepsilon^{(3)}_\alpha&=&\frac{1}{\eta^2\sqrt{k^2}}
\left(\frac{\sqrt{a}}{\sqrt{\beta}}\eta^2 k_{\alpha}+
\frac{\sqrt{a}}{\sqrt{\beta}}(FFk)_{\alpha}-
\frac{\sqrt{\beta}}{\sqrt{a}}(FFk)_{\alpha}\right),\nonumber
\ea
where the following notations are adopted
\[
a=kFFk,\,\, \beta=a+\eta^2 k^2,\,\,
\eta=\left(\frac{1}{2}F_{\mu\nu}F^{\mu\nu}\right)^{1/2},
\]
the first two of them being for the transversal photon
polarization, and the third for the longitudinal one.

As a result, upon assuming without any loss of generality that
${\bf k} = (0,0,|{\bf k}|),$ ${\bf H} = (H, 0, 0),$ the width of the virtual
photon decay into a neutrino--antineutrino pair $\Gamma(i)$ ($i$
``enumerates'' the photon polarization, $i=1,2,3$) is obtained
\ba \label{total}
\Gamma(1)&=&\frac{e^2G_F^2}{3(4\pi)^5} \cdot k^2 \langle A_1^a \rangle _a
\langle A_1^{a'+} \rangle _{a'}, \nonumber\\
\Gamma(2)&=&\frac{e^2G_F^2}{3(4\pi)^5} \cdot \Bigl[ k^2 \langle A_1^a
\rangle _a \langle A_1^{a'+} \rangle _{a'}+{\bf k}{}^2\langle B^a \rangle _a
\langle B^{a'+} \rangle _{a'} \Bigr], \nonumber\\
\Gamma(3)&=&\frac{e^2G_F^2}{3(4\pi)^5} \cdot \Bigl[{\bf k} {}^2
\langle B^a \rangle _a \langle B^{a'+} \rangle _{a'} + \left| k_0
\langle A_3^a \rangle _a  - |{\bf k}| \langle D^a \rangle _a \right|^2 \Bigr],
\ea
where
\ba \label{very}
A_1^a &=& {\ds \int\limits_{-1}^1 dv \Bigl\{ \frac{1-v^2}{2}
f_1(\tilde z_a) k^2 + \frac{(1/3 - v^2)b^2z_a}{2m_a^2}
[z_a f'(\tilde z_a) - f(\tilde z_a)]} - \nonumber\\
&&- {\ds\frac{(1-v^2)b^2z_a^2}{24m_a^4\eta^2}f'(\tilde z_a)
[2(2-v^2)a + (3v^2 - 1)\beta]} \Bigr\},\nonumber\\
A_2^a &=& {\ds \int\limits_{-1}^1 dv \Bigl\{ \frac{1-v^2}{2}
f_1(\tilde z_a) k^2 - \frac{b^2z_a}{3m_a^2}
[z_a f'(\tilde z_a) + f(\tilde z_a)]} - \nonumber\\
&&- {\ds \frac{(1-v^2)b^2z_a^2}{12m_a^4\eta^2}f'(\tilde z_a)
[(2-v^2)a + \beta]} \Bigr\}, \nonumber\\
A_3^a &=& {\ds \frac{\sqrt \beta}{\eta \sqrt{k^2}}
\int\limits_{-1}^1 dv \Bigl\{ \frac{1-v^2}{2} f_1(\tilde z_a)
\frac{\beta + \eta^2 k^2}{2\eta^2} - \frac{b^2z_a}{2m_a^2} (v^2 -1/3)\times}
\nonumber\\
&&\times{\ds [z_a f'(\tilde z_a) - f(\tilde z_a)]
- \frac{(1-v^2)b^2z_a^2}{48m_a^4\eta^2}f'(\tilde z_a)
[\beta (v^2 - 3) + \eta^2 k^2 (1+5v^2)]} \Bigr\},\nonumber\\
D^a &=& -{\ds \frac{\sqrt a}{\eta \sqrt{k^2}}\int\limits_{-1}^1 dv
\Bigl\{ \frac{1-v^2}{2} f_1(\tilde z_a)
\frac{a - \eta^2 k^2}{2\eta^2} + \frac{b^2z_a}{3m_a^2}
[z_a f'(\tilde z_a) + f(\tilde z_a)]} + \nonumber\\
&&+ {\ds \frac{(1-v^2)b^2z_a^2}{48m_a^4\eta^2}f'(\tilde z_a)
[a (3 - v^2) + \eta^2 k^2 (3v^2 - 5)]} \Bigr\}, \nonumber\\
B&=&{\ds\int\limits_{-1}^1dv\, b z_a f(\tilde z_a)},\,\,\,
C = {\ds\int\limits_{-1}^1dv\, \frac{1-v^2}{2} \frac{bz_a}{2m_a^2}},\,\,\,
b=e\eta.\nonumber
\ea
In this expression the Airy--Hardy functions
\[ \label{H}
f(\tilde z_a)= \Upsilon(\tilde z_a) +i \Phi (\tilde z_a) =
i \int\limits_0^\infty ds\,exp[-i(s \tilde z_a+\frac{s^3}{3})],
\]
are introduced, as well as their derivatives $f'(\tilde z_a)$ and a function
\[
f_1(z) = \int\limits_0^\infty \frac{du}{u} e^{-izu}
\left( e^{-iu^3/3} - 1\right)
\]
The argument of these functions has the form
\[ \label{tz}
\tilde z_a = \left( 1- \frac{1-v^2}{4} \frac{k^2}{m_a^2}\right)
z_a, \;\; z_a =  \left( \frac{4}{\chi_a(1-v^2)}\right)^{2/3},
\]
where an invariant dynamic parameter
\[
\chi_a = \frac{e\sqrt{kF^2k}}{m_a^3} = \frac{eH|{\bf k}|}{m_a^3}
\]
was introduced.

In the limiting case when a low virtuality photon decays, i.e.,
$k^2 = m_\gamma^2 > 0$ $(k_0^2\gg {\bf k} {}^2)$ and
$m_\gamma^2 \ll m_a^2$, Eq. (2) yields
\ba
\Gamma(1)&=&\frac{e^2G_F^2}{3(4\pi)^5} m_\gamma^2
\langle \frac{4}{45} \frac{e^2 H^2}{m_a^2}
\frac{m_\gamma^2}{m_a^2} \rangle_a^2, \nonumber\\
\Gamma(2)&=&\frac{e^2G_F^2}{3(4\pi)^5} m_\gamma^2
\left[\langle \frac{2}{9} \frac{e^2 H^2}{m_a^2}
\frac{m_\gamma^2}{m_a^2} \rangle_a^2 + \frac{{\bf k} {}^2}{m_\gamma^2}
\langle 2eH(1+\frac{m_\gamma^2}{6m_a^2}) \rangle_a^2 \right], \nonumber\\
\Gamma(3)&=&\frac{e^2G_F^2}{3(4\pi)^5} m_\gamma^2
\left[\langle \frac{4}{45} \frac{e^2H^2}{m_a^2}
\frac{m_\gamma^2}{m_a^2} \rangle_a^2 + \frac{{\bf k}{}^2}{m_\gamma^2}
\langle 2e H(1+\frac{m_\gamma^2}{6m_a^2}) \rangle_a^2 \right].\nonumber
\ea

     Next asymptotics of the rate are presented which describe a
process that corresponds to the scattering in an external field
channel, when $k^2 = - m_\gamma ^2 <0$. In the case of scattering
on a high virtuality photon $( m_\gamma^2 \gg m_a^2)$, when
${\bf k}{}^2 \gg k_0^2$ the rate of the process is described as
\ba \label{a}
w(1)&=&\frac{e^2G_F^2}{3(4\pi)^5 k_0} m_\gamma^2 \langle \frac{8e^2H^2}{3m_\gamma^2} \ln{\frac{m_a^2}{m_\gamma^2}} \rangle^2_a, \nonumber\\
w(2)&=&w(3) = \frac{e^2G_F^2}{3(4\pi)^5 k_0} m_\gamma^2
\langle \frac{4e Hm_a^2}{m_\gamma^2} \ln{\frac{m_a^2}{m_\gamma^2}} \rangle^2_a,
\ea
and for scattering on a low virtuality photon $(m_\gamma ^2 \ll m_a^2)$
and ${\bf k}{}^2 \gg k_0^2$ we have
\ba \label{b}
w(1)&=&\frac{e^2G_F^2}{3(4\pi)^5 k_0} m_\gamma^2
\langle \frac{4}{15} \frac{e^2 H^2m_\gamma^2}{m_a^4} \rangle^2_a,\\
w(2)&=&w(3) = \frac{e^2G_F^2}{3(4\pi)^5 k_0} m_\gamma^2 4 e^2 H^2.\nonumber
\ea

     According to the remarks of Refs. [4, 6], the influence of an
external field results in the fact that all the self--energy corrections
to an external (real) photon line in an external field account for the
appearance of a finite mass of a photon. Thus, formulas (3), (4) obtained
above may be applied for making estimates of the real photon life time with
respect to the decay into a neutrino--antineutrino pair in a magnetic field.



\end{document}